\documentclass[9pt,twocolumn,twoside,lineno]{pnas-new}

\usepackage{siunitx}
\usepackage{amssymb}

\newcommand{\diag}[1]{\mathrm{diag}[#1]}
\newcommand{\bfg}{\mathbf{g}}
\newcommand{\bphi}{\boldsymbol{\varphi}}
\newcommand{\bfu}{\mathbf{f}}
\newcommand{\bfH}{\mathbf{H}}
\newcommand{\bu}{\mathbf{u}}
\newcommand{\sst}[2]{{#1}_{\text{#2}}}

\templatetype{pnasresearcharticle} 

\title{High-Resolution Limited-Angle Phase Tomography of Dense Layered Objects Using Deep Neural Networks}

\author[a,1]{Alexandre Goy}
\author[b]{Girish Rughoobur} 
\author[a]{Shuai Li}
\author[a]{Kwabena Arthur}
\author[b]{Akintunde I. Akinwande} 
\author[a,c]{George Barbastathis}

\affil[a]{Mechanical Engineering, Massachusetts Institute of Technology, Cambridge, MA02139, USA.}
\affil[b]{Electrical Engineering and Computer Science, Massachusetts Institute of Technology, Cambridge, MA02139, USA.}
\affil[c]{Singapore-MIT Alliance for Research and Technology (SMART) Centre, Singapore 117543, Singapore.}

\leadauthor{Goy}

\significancestatement{The authors demonstrate that it is possible to use Deep Neural Networks to produce tomographic reconstructions of dense layered objects with small illumination angle as low as 10$^\circ$. It is also shown that a DNN trained on synthetic data can generalize well to and produce reconstructions from experimental measurements. This work has application in the field of X-ray tomography for the inspection of integrated circuits, and other materials studies.}

\authorcontributions{A.G. performed the experiments, performed the numerical simulations and wrote the manuscript, G.R. fabricated the samples, S.L. and K.A. contributed to the DNN design.}
\authordeclaration{The authors declare no conflict of interest.}
\correspondingauthor{\textsuperscript{1}To whom correspondence should be addressed. E-mail: agoy@mit.edu}


\begin{abstract}
We present a Machine Learning-based method for tomographic reconstruction of dense layered objects, with range of projection angles limited to $\pm $10$^\circ$. Whereas previous approaches to phase tomography generally require two steps, first to retrieve phase projections from intensity projections and then perform tomographic reconstruction on the retrieved phase projections, in our work  a physics-informed pre-processor followed by a Deep Neural Network (DNN) conduct the three-dimensional reconstruction directly from the intensity projections. We demonstrate this single-step method experimentally in the visible optical domain on a scaled up integrated circuit phantom. We show that even under conditions of highly attenuated photon fluxes a DNN trained only on synthetic data can be used to successfully reconstruct physical samples disjoint from the synthetic training set. Thus, the need of producing a large number of physical examples for training is ameliorated. The method is generally applicable to tomography with electromagnetic or other types of radiation at all bands. 
\end{abstract}

\dates{This manuscript was compiled on \today}
\doi{\url{www.pnas.org/cgi/doi/10.1073/pnas.XXXXXXXXXX}}

\begin{document}

\maketitle
\thispagestyle{firststyle}
\ifthenelse{\boolean{shortarticle}}{\ifthenelse{\boolean{singlecolumn}}{\abscontentformatted}{\abscontent}}{}

\dropcap{T}omography is the quintessential inverse problem. Since the interior of a three-dimensional object is not accessible non-invasively, the original insight of tomographic approaches was to illuminate through from multiple angles of incidence and then process the resulting projections to reconstruct the interior slice-by-slice\ \cite{inv:bracewell54,inv:bracewell56,inv:bracewell56b}. In the simplest case, when diffraction is negligible and the illumination is collimated, as is generally permissible to assume for X-ray attenuation\ \cite{inv:cormack63,inv:cormack64,inv:hounsfield73,inv:ambrose73} and electron scattering\ \cite{inv:derosier68,inv:derosier70,inv:crowther70} in the far field and for features of size $\sim$\SI{1}{\micro\meter} and above, the object's interior is represented by its Radon transform\ \cite{inv:radon1917} of line integrals along straight parallel paths. The interior of the volume is then reconstructed by use of the Fourier-slice theorem for the Radon projections. On the other hand, if the X-ray beam is not collimated but spherical, then the slice-by-slice approach is no longer applicable and full volumetric reconstruction is required \cite{horn:1978, horn:1979}. Even when the object is available for observation from the full $360^\circ$ range of projection angles, these instances of tomography are all highly ill-posed because the Fourier-slice property results in uneven coverage of the Fourier space with the high spatial frequencies ending up underrepresented. Ill-posedness increases when the angular range is limited because then an entire cone of spatial frequencies goes missing from the measurement. Alternatively, in this case, Tomosynthesis\ \cite{inv:dobbins03} utilizes sheared (rather than rotated) projections to bring slices from within the interior into focus, but with lower contrast since emission from the rest of the volume remains as background. 

Additional challenges occur when the inverse problem of interest is to reconstruct in 3D the index of refraction, rather than the attenuation. If the object features are large enough compared to the wavelength, such that diffraction may still be neglected, and the index variations through the object volume are relatively small, then each projection may be modeled as a set of Fermat integrals of phase delay along approximately straight lines. The phase integrals may be obtained, for example, using holographic interferometry\ \cite{vest:1972, vest:1979} or transport of intensity \cite{petruccelli:2013}. For smaller-sized features and still assuming weak scattering ($1^{\text{st}}$-order Born approximation), the projection integrals are instead obtained along curved paths on the surface of the Ewald sphere, a method referred to as diffraction tomography \cite{wolf:1969, choi:2007}. By decoupling the problem into two parts: first phase projection retrieval, followed by tomography, these approaches enjoy the benefit of using the advanced algorithms in the two respective research fields. However, there is also the danger that errors generated independently during each step may amplify each other. Lastly, when strong scattering may no longer be neglected, all two-step approaches become questionable because the interpretation of the first step as line integrals is no longer valid. 

Generally, ill-posed inverse problems are solved by regularized optimization. If $f$ is the object and $g$ the measurement, then the object estimate $\hat{f}$ is obtained as\ \cite{inv:tikhonov63a, inv:tikhonov65}
\begin{equation}
\hat{f}=\underset{f}{\operatorname{argmin}} \left\{\| Hf-g \|^2 + \alpha \Phi(f)
\right\}.
\label{eq:tw}
\end{equation}
Here, $H$ is the forward operator relating the measurement to the object, $\Phi$ is the regularizer expressing prior knowledge about the object, and $\alpha$ is the regularization parameter controlling the competition between the two terms. The prior may be thought of as rejecting solutions to the inverse problem that are known to violate known properties of the class of objects being imaged; for example, if the class where $f$ belongs is known to have sharp edges, then the regularizer should be applying a high penalty to blurry solutions $\hat{f}$. Thus, the inherent uncertainty due to ill-posedness is reduced. Sparsity-promoting compressive priors\ \cite{inv:candes-tao05,inv:candes-romberg-tao06a,inv:donoho2006compressed,inv:candes-romberg-tao06b} found some of their first successes in tomographic reconstruction\ \cite{inv:candes-tao06,inv:brady15-compressive-tomo}. Compressive sensing is directly implemented through a proximal gradient solution to Eq.~\ref{eq:tw} if a set of basis functions where the object class is sparse is {\it a priori} known. Alternatively, if a database of representative objects is available, then these examples may be used to {\em learn} the optimal set of basis functions as a dictionary\ \cite{inv:elad2006image,inv:aharon2006k}. 

Rapid recent developments in the field of Machine Learning, and Deep Neural Networks\ \cite{nn:lecun15-dl} (DNNs) in particular, have provided an entirely novel set of tools and insights for inverse problems. It may be shown\ \cite{inv:Gregor10,mardani:2017} that recurrent or unfolded multi-stage DNN architectures are formally equivalent to the iterative solution to the inverse problem in Eq.~\ref{eq:tw} where the prior $\Phi$ need no longer be known or depend on sparsity; instead, examples guide the discovery of the prior through the DNN training process. Simpler learning architectures, where $g$ is fed to the DNN directly or after first passing through a pre-processor, have been used for retrieval of phase from intensity\ \cite{inv:sinha17-PhENN,inv:sinha17-PhENN,rivenson:2018,inv:PhENN-spectral-premod,kemp:2018}; 3D holographic reconstruction\ \cite{inv:shimobaba18,inv:ren18-dh-autofocus-dl,inv:ozcan-dnn-extDOF}; super-resolution photography\ \cite{inv:dong14-super-res,inv:dong15-super-res,inv:perceptual-loss} and microscopy\ \cite{inv:rivernson17-dlm}; imaging through scatter\ \cite{inv:Lyu2017,inv:IDiffNet,inv:Borhani18,inv:Li18-deep-speckle}; and imaging under extremely low light conditions in the three contexts: computational ghost imaging\ \cite{inv:Lyu2017b}, consumer-camera photography\ \cite{inv:chenchen18-dark}, and phase retrieval\ \cite{goy:2018}.


Multi-stage DNN architectures have been shown to yield high-quality reconstructions in numerous Radon tomography configurations\ \cite{mardani:2017,inv:mardani2017a,schlemper:2017, jin:2017, inv:McCann2017, gupta:2018, sun:2018}. Recently, Nguyen \textit{et al.} \cite{nguyen:2018} used the inverse Radon transform for optical tomography with a single-stage DNN intended to partially correct for the assumption of line integrals breaking down. 

In this paper, we apply a Fourier based beam propagation method (BPM) \cite{feit:1978} as a pre-processing step immune from any Radon assumptions. The strongly scattering object is illuminated by a parallel beam under a limited angular range of $20^\circ$, {\it i.e.} $\pm 10^\circ$ from the reference axis. Unlike the earlier works on refractive index tomography referenced above, we do not perform a phase retrieval step; rather, the intensity measurements are pre-processed to produce directly an initial crude three-dimensional guess of the object's interior. This crude guess is then fed to our machine learning algorithm. The pre-processing step is necessary because, even if we did convert intensity to phase, the results would not be interpretable as line integrals under our experimental conditions. Moreover, by merging phase retrieval and tomography into a single step, our algorithm becomes less sensitive to error accrual. 

Large data sets, typically consisting of more than 5,000 examples, are generally required for DNN training. That is feasible in many cases through spatial light modulators\ \cite{inv:sinha17-PhENN} or\ \cite{nguyen:2018}. However, in many cases of interest spatial light modulators have insufficient space-bandwidth product or are unavailable (e.g. in X-rays); and alternatives to generate physical specimens are expensive or restricted due to proprietary processes. Instead, our approach is to train the DNN on purely synthetic data  with the rigorous BPM forward model, and then use a physical test specimen (phantom) to test the reconstruction quality with well calibrated ground truth in experiments. 

We chose to design our phantom as emulating the three-dimensional geometry of integrated circuits (ICs). These would normally be inspected with X-rays, so we scaled-up the feature dimensions in the phantom for visible wavelengths. The advantage of this choice is that IC layouts provide strong geometrical priors, e.g. Manhattan geometries, and our phantom also exhibited large spatial gradients and refractive index contrast to strengthen scattering. Thus, our methodology is directly applicable to all cases of tomography at optical wavelengths, e.g. 3D-printed specimen characterization and identification, and biological studies in cells and tissue with moderate scattering properties. In each case, testing the ground truth would require the fabrication or the accurate simulation of different phantoms meeting the corresponding priors. 

There is also value in the study of emulating X-ray inspection of ICs at visible wavelengths, as extensive outsourcing by the IC industry has created a growing concern that the ICs delivered to the customer may not match the expected design, and that malicious features may have been added\ \cite{mak:2015}. However, in our emulation the phase contrast of the features against the background and the Fresnel number are both higher than typical corresponding IC configurations even at soft X-ray wavelengths.

One advantage of deep learning for inverse problems is speed. Solving (\ref{eq:tw}) separately for each instance of $g$ is computationally intensive, and training a DNN is even more so. This is because both operations are iterative, and the latter is run on large datasets. On the other hand, once the DNN has learnt the inverse map from the pre-processed version of $g$ to $\hat{f}$, the computations are feedforward only. For example, the IC layout priors we exploit here could, in principle, also be learnt by dictionaries---but, under strong scattering conditions, the latter would require iterative optimization of\ Eq.~\ref{eq:tw} with the forward operator $H$ itself consisting of an expensive computational procedure in each iteration. In our approach, the pre-processing performed prior to the DNN is the most time consuming operation, therefore we aim at simplifying the pre-processing step as much as we can, \textit{i.e.} tolerate a crude approximation, and leave it to the DNN to correct it. In our case, execution times is 51~sec (out of which only 300msec are taken by the DNN, the rest is the pre-processor) while and Learning Tomography\ \cite{kamilov:2015, kamilov:2016}, which is based on a similar gradient descent algorithm, takes 212~sec and yields inferior reconstructions (see Fig.~\ref{fig:reconstructions} in the Results section).

\section*{Optical experiment}

\begin{figure}[ht!]
  \begin{center}
    \includegraphics{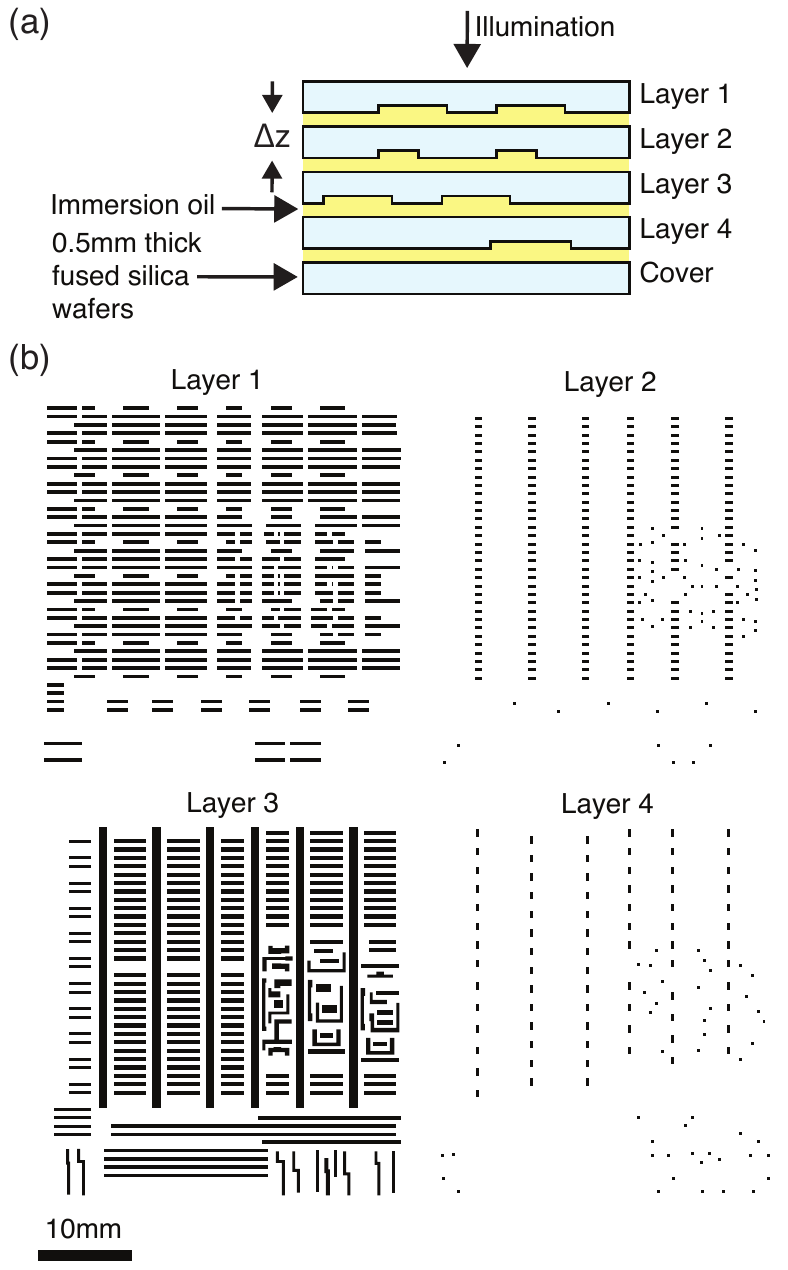}
     \caption{\small{\textbf{(a) Sample cross section.} The depth of the etched patterns was measured (see Materials and Methods section) and the refractive index of the oil was controlled in order to achieve a known phase shift of -0.32rad. $\Delta z = 0.5$mm. \textbf{(b) IC patterns} used for each of the four layers. The white background represents the original wafer thickness and the black areas indicates where the wafer has been etched.}}
	 \label{fig:sample}
  \end{center}
\end{figure}

We prepared a series of four glass wafers with etched structures representing patterned layers from an actual IC design. A schematic cross-section of the sample is shown in Fig.~\ref{fig:sample}a. The glass plates are held together and aligned in a custom made holder. Immersion oil is added between the plates to minimize parasitic reflections and also tune the phase shift associated with the patterns. The pattern depth was measured to be 575nm, yielding a phase shift of $-0.33$rad for the particular oil used. Note that the phase shift is negative as the refractive index of the oil is lower than the refractive index of the glass. Details about the sample preparation and phase shift measurements are given in the Materials and Methods section. The particular patterns etched on the sample are shown in Fig.~\ref{fig:sample}b.

\begin{figure}[ht]
  \begin{center}
    \includegraphics{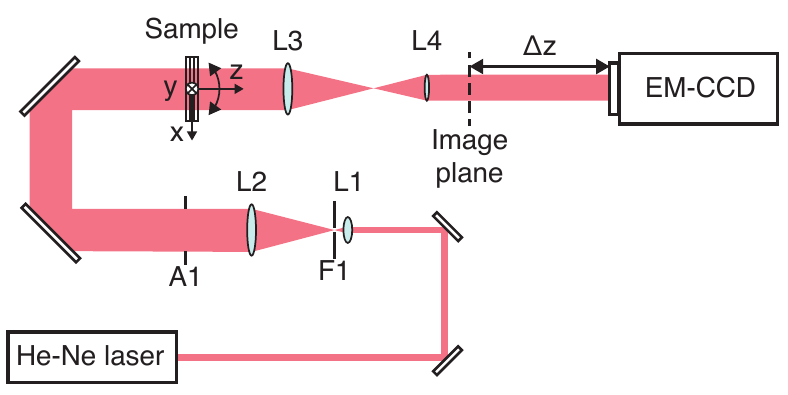}
     \caption{\small{\textbf{Experimental apparatus.} Spatial filter and beam expander: L1 is 10$\times$, 0.25 NA objective, L2: $\SI{100}{\milli\meter}$ lens, with a $\SI{5}{\micro\meter}$ pinhole F1 in the focal plane, L3: $\SI{200}{\milli\meter}$ lens, L4: $\SI{100}{\milli\meter}$ lens. Aperture A1 cuts the outer diffraction lobes of the beam. The sample is mounted on a 2-axis rotation stage rotating along the $x$ and $y$ axes. The sample middle plane is imaged using a telescope lens system with magnification $0.50\times$. The camera is defocused by a distance $\Delta z =$ $\SI{58}{\milli\meter}$ from the image plane.}}
	 \label{fig:setup}
  \end{center}
\end{figure}

The experimental apparatus is detailed in Fig.~\ref{fig:setup}. A collimated monochromatic plane wave from a CW laser is incident on the sample, which is mounted on a two-axis rotation stage. The sample is imaged through a demagnifying telescope to increase the field of view. The detector (an EM-CCD camera) is defocused from the image plane to simulate free space propagation in an X-ray experiment where no imaging system can be used. Further details are given in the Materials and Methods section. 

The strength of diffraction effects can be quantified with the Fresnel number $F = a^2/(\lambda d)$, where $\lambda$ is the wavelength, $d$ the propagation distance and $a$ the characteristic feature size of the object. The smaller the Fresnel number, the stronger the effects of diffraction. For the glass phantom considered here and a defocus of 58mm, $F = 0.7$ for the smallest features and $F = 5.5$ for the largest. The diffraction pattern is digitized on the camera for different sample orientations. This series of measurements is then passed through a numerical algorithm, described in the next section, whose aim is to yield a first approximate reconstruction, hereafter referred to as the ``Approximant.''

\section*{Computation of the Approximant}

As mentioned above, the task of the DNN is significantly facilitated if the raw measurements are pre-processed so as to give an approximation of the solution. We use a simple gradient descent method to generate the approximant of the sample refractive index distribution.

Light propagation through the object can be computed using the following split-step Fourier BPM. In this method each sample layer $l$ is modeled as a two-dimensional complex mask $f_l(x, y) = \exp[\alpha_l(x, y) + j\varphi_l(x, y)]$ and the space between two successive layers as  Fresnel propagation  through a homogeneous medium whose index of refraction equals the average refractive index of the sample, as
\begin{equation}\label{equ:bpm}
		u_l = \mathcal{F}^{-1}\left\{\mathcal{F}\left\{u_{l-1}f_{l-1}\right\}(k_x, k_y) e^{-j\left(k-\sqrt{k^2 - k_x^2 - k_y^2}\right)\Delta z}.\right\}
\end{equation}
Here, $u_l(x, y)$ is the optical field at layer $l$, $\mathcal{F}$ the Fourier transform, $\Delta z$ the distance between layers and $k$ the wavenumber in the medium between layers. Each measurement is a collection of $N_v$ intensity patterns $g_i(x, y)$, with $i = 1,...,N_v$, captured on the detector for each orientation $i$ of the sample. In this work, we assume that the sample is a pure phase object, \textit{i.e.} $\alpha(x, y) = 0$. This assumption is valid for the glass phantom.

From the measurements, we produce an approximation $\tilde{\mathbf{f}}$ of the phase pattern $\varphi_l(x, y)$ for each layer $l$ in the sample. We use the steepest gradient descent method with a fixed number of iterations $K=8$ to generate the Approximant. In what follows, we represent the measurements (consisting of $M$ real pixel values) by a ($M\times 1$) column vector $\bfg_i$ and the discretized object (consisting of $N$ real voxel values) by a ($N\times 1$) column vector $\bfu$. We then define a cost function $J$ to minimize, consisting simply of a data fidelity term:
\begin{equation}\label{equ:problem}
J = \frac{1}{2}\sum_{i=1}^{N_v}\|\sst{H}{i}(\bfu) - \bfg_i\|^2_2,
\end{equation}
where $\sst{H}{i}$ denotes the forward operator that maps the object function $\bfu$ to a prediction of the measurement $\sst{H}{i}(\bfu)$, for a particular orientation $i$ of the sample. In the problem presented here, the optical field will first propagate through the sample $L$ layers, each of thickness $\Delta z$, and then in free space to the detector over a distance $d$. The forward operator can thus be written as a cascade of Fresnel propagation operations and thin mask multiplications corresponding to the object layers, \textit{i.e.} successive applications of Eq.~\ref{equ:bpm} written in operator form:
\begin{eqnarray}\label{equ:forwardop0}
\sst{H}{i}(\bfu) &=& \left|\bu_{\mathrm{det}}\right|^2\\\label{equ:forwardop01}
\bu_{\mathrm{det}} &=& F_d \diag{\bfu_L}...F_{\Delta z}\diag{\bfu_2}F_{\Delta z}\diag{\bfu_1}\bu_{\mathrm{inc}, i},
\end{eqnarray}
where $\bfu_l$ is the vector of object function values in layer $l$, $F_{\Delta z}$ the Fresnel propagation operator over distance $\Delta z$, $\bu_{\mathrm{inc}, i}$ the incident field, $\bu_{\mathrm{det}}$ the field on the detector, and $\diag{\mathbf{v}}$ the diagonal matrix with vector $\mathbf{v}$ on the diagonal. The gradient descent iterative update can be written as
\begin{equation}\label{equ:grad_desc}
\bfu^{(k+1)} = \bfu^{(k)} - s(\nabla_{\bfu^{(k)}} J)^T,
\end{equation}
where $\bfu^{(k)}$ is the object estimate at iteration $k$, $s$ the step size, and $\nabla_{\bfu^{(k)}}J$ the gradient of $J$ with respect to $\bfu$ evaluated at $\bfu^{(k)}$. We then set $\tilde{\bfu} = \bfu^{(K)}$ with $K$ chosen in advance, starting from $\bfu^{(0)} = \mathbf{0}$. The detailed derivation of the gradient for the particular model in Eqs.~\ref{equ:forwardop0}-\ref{equ:forwardop01} is given in the appendix. In Fig.~\ref{fig:measurements}, we give an example of one experimental intensity measurement ($\bfg_1$) taken from the series of tomographic projections and the corresponding approximant $\bfu^{(8)}$ obtained from the whole series using Eq.~\ref{equ:grad_desc}.

\begin{figure}[ht!]
  \begin{center}
    \includegraphics{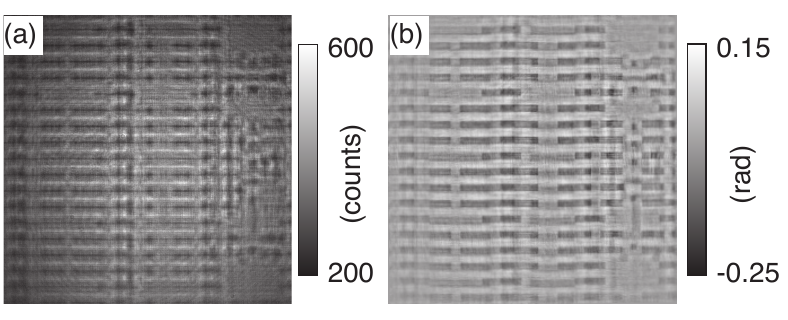}
		 \caption{\small{(a) Examples of experimental intensity measurement for the following sample orientation $(\theta_x = -10^\circ, \theta_y = 0^\circ)$. (b) Phase approximant for IC layer 1 obtained from the collection of 22 intensity patterns at different orientations $(\theta_x = -10^\circ, -8^\circ, ..., +10^\circ, \theta_y = 0^\circ)$ and $(\theta_x = 0^\circ, \theta_y = -10^\circ, -8^\circ, ..., +10^\circ)$.}}
	 \label{fig:measurements}
  \end{center}
\end{figure}

\section*{DNN architecture and training}

We use a DNN to map the approximant to the final reconstruction $\hat{\bfu}$. The DNN is a convolutional neural network with a DenseNet architecture \cite{huang:2016}. The implementation is the same as the DNN used in \cite{inv:IDiffNet} except that the number of dense blocks was reduced to 3 in both the encoder and the decoder, as we empirically observed that using more dense blocks did not result in a significant improvement of the results. We produce a total of 5,500 synthetic sets of measurements obtained by simulating the optical apparatus with the beam propagation method in Eq.~\ref{equ:bpm}. The synthetic measurements were subject to simulated shot noise and read noise equivalent to the noise levels found in the experiment. The shot noise was accounted for by converting the simulated measurement pixel intensities $I$ expressed in average photon count per pixel per integration period of the detector to integer numbers of photons $N$ following Poisson statistics. 
The actual optical power on the camera was measured with a power meter and converted to an average photon flux per detector pixel. Read noise following a Gaussian statistics was subsequently added. The parameters (variance and average) of the noise were measured from a series of dark frames from the camera taken with the same gain (EM gain of 1) and integration time (2ms) than the experimental measurements.

From each set of measurements, we produce a multi-layer Approximant using the gradient descent in Eq.~\ref{equ:grad_desc}. The examples are split in a training set of 5,000 examples, a validation set of 450 examples and a test set of 50 examples. Each set of measurements (one example) comprises 22 views corresponding to different orientations of the sample. The DNN is then trained to map the Approximant to the ground truth used for the simulation. Each layer of the sample is assigned to a different channel in the DNN. We use the negative Pearson correlation coefficient ($\text{NPCC} = -\text{PCC}$) as loss function and train in 20 epochs with a batch size of 16 examples. For two images $A$ and $B$, the PCC is defined as:
\begin{equation}\label{equ:npcc}
    \mathrm{PCC}(A, B) = \frac{\sum_{i}(A_{i}-\bar{A})(B_{i}-\bar{B})}{\sqrt{\sum_{i}(A_{i}-\bar{A})^2\sum_{i}(B_{i}-\bar{B})^2}},\\
\end{equation}

The PCC (and therefore the NPCC too) is agnostic to scale and offset, \textit{i.e.} $\mathrm{PCC}(aA + b, cB + d) = \mathrm{PCC}(A, B)$ for $a, b, c, d \in \mathbb{R}$. As a consequence, the DNN, which is trained by minimizing the NPCC, may apply some offset and scaling to the reconstruction. These parameters are not easily predictable; however, for a given DNN they are constant once training is complete, which allows us to correct the reconstructions. Offset and scale are obtained by least square linear regression between the DNN output and the ground truth from the synthetic test set examples (not including the experimental example).

\section*{Results}

\begin{figure*}[ht!]
  \begin{center}
    \includegraphics{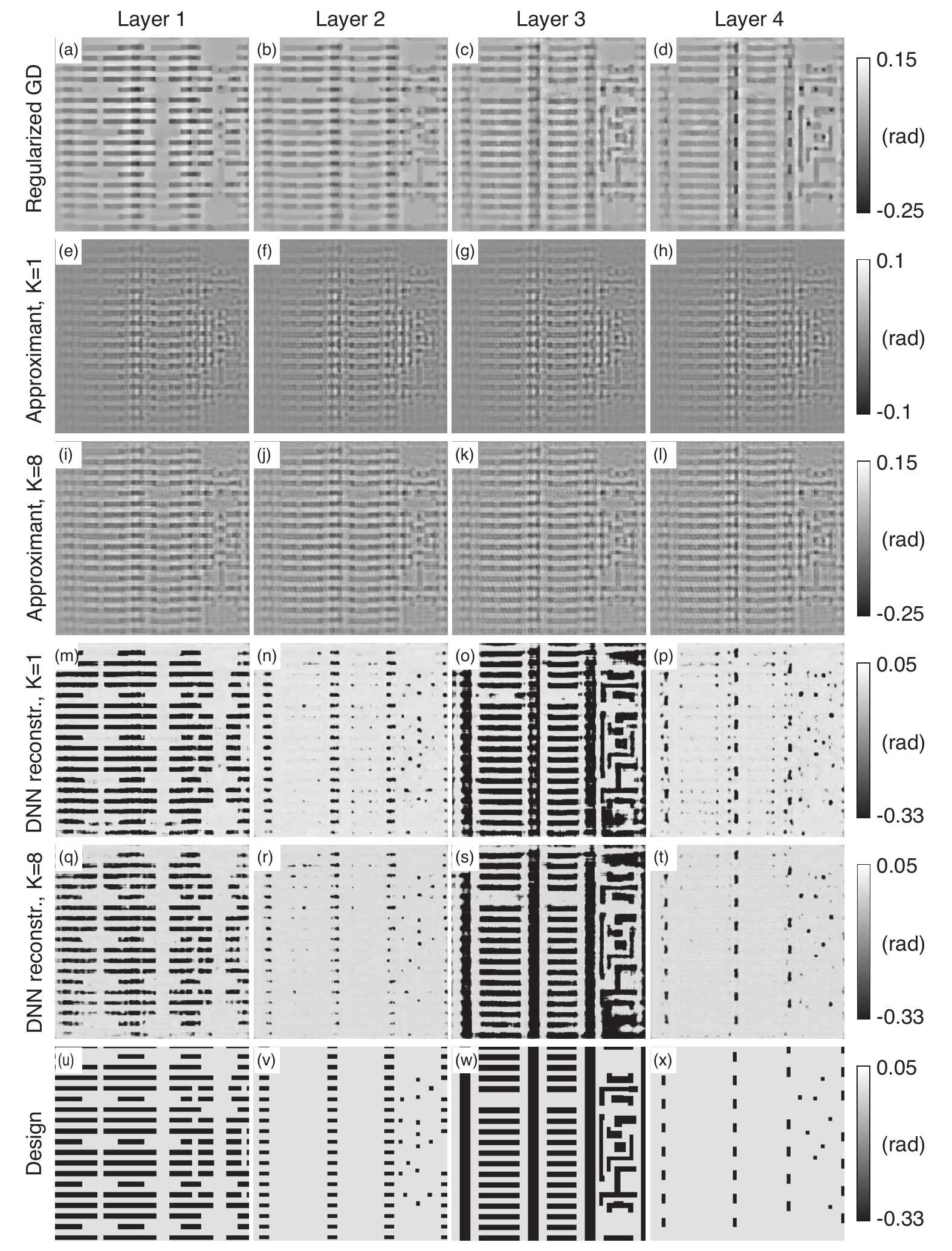}
     \caption{\small{(a) to (d) Proximal gradient descent with TV regularization, $K=30$ iterations, for each layer 1 to 4. (e) to (h) Approximants generated from the experimental measurements with $K=1$. (i) to (l) Approximants generated from the experimental measurements with $K=8$. (m) to (p) Reconstructions from the DNN of each approximant e to h respectively. (q) to (t) Reconstructions from the DNN of each approximant i to l respectively. (u) to (x) Idealized ground truth obtained from the sample specifications for layer 1 to 4. Note that the color bar range covers more than the range of the data, there is no saturation effect on the images.}}
	 \label{fig:reconstructions}
  \end{center}
\end{figure*}

The method described in the previous sections was applied to the glass phantom shown in Fig.~\ref{fig:setup}c. The synthetic measurements were subject to Poisson noise resulting from $10^3$ photon flux per detector pixel, equal to the experimental photon flux, and an additive Gaussian noise with a standard deviation of 13 counts. For DNN training, we compared two sets of approximants, obtained with $K=1$ and $K=8$ with and wihtout TV regularization. In the case $K=1$, the regularization parameter $\alpha$ was set to 0.1 (step size 0.1). We chose a smaller value of 0.04 for the case $K=8$ (step size 0.05) because the proximal operator corresponding to the regularizer is applied at each iteration and its effect tends to accumulate. In the case $K=8$, the particular choice for the number of iterations is an empirical trade-off between computation time and accuracy. The same optimization parameters (step size and number of iterations) were used to compute the approximant of the IC phantom, the result for each layer is shown in Fig.~\ref{fig:reconstructions}e to h for $K=1$ and Fig.~\ref{fig:reconstructions}i to l for $K=8$. The DNN reconstruction results are summarized in Fig.~\ref{fig:reconstructions}m to p ($K=1$) and q to t ($K=8$). The approximant and the DNN reconstruction represent the phase modulation imposed by each layer in the sample. The absolute phase carries no useful information, therefore we are free to offset the reconstructed phase by an arbitrary constant. In the DNN reconstructions in Fig.~\ref{fig:reconstructions}i to l, the IC patterns (where the phase shift actually occurs) are typically reconstructed with zero phase due to the rectified linear units (which project all negative values to 0) at the output layer of the DNN. We reassign the phase of the pattern to the nominal phase of -0.33rad so that it can be visually compared to the ground truths in Fig.~\ref{fig:reconstructions}m to p. An alternate approach leading to very similar results is to assign a zero phase to the background.

The DNN reconstructions can be compared to those obtained using Learning Tomography (LT), a previously demonstrated optical tomography technique \cite{kamilov:2015, kamilov:2016} based on proximal optimization (FISTA) \cite{beck:2009} with total variation (TV) regularization \cite{beck:2009b}. The role of the TV regularizer is to favor piecewise constant solutions while preserving sharp edges, which is especially well suited for IC patterns. LT was initially designed for holographic measurements and was modified here to work on intensity measurements by computing the gradient for the data fidelity term in Eq.~\ref{equ:problem}. The essential difference in the LT optimization is that a TV filter playing the role of a proximal operator is applied at each iteration on the current solution. The LT reconstructions for the experimental data set are shown in Fig.~\ref{fig:reconstructions}a to d. These particular reconstructions were obtained after 30 iterations of gradient descent, a step size of 0.05, a regularization parameter $\alpha = 0.04$ and 20 iterations for the TV regularizer at each step. The computation time of the $\bfu^{(8)}$ approximant is 51~sec for $K=8$ (no regularization) and 6~sec for $K=1$ on an Intel i9-7900X processor running at 3.3~GHz, including 570~msec for the DNN run on an NVIDIA Titan Xp graphics processing unit, \text{vs.} 212~sec for LT on the same processor.

In Table~\ref{tab:pcc}, we summarize the values of the PCC, which we use to quantify the quality of the reconstructions. The values are given for reconstructions on the synthetic test set (50 examples) and also the reconstruction of the single experimental example. Because the reconstruction quality turns out to depend strongly on the particular layer, we display the value for each four layers separately. As may be expected, the values for the LT are higher (better reconstructions) than those for the approximant as LT was run for 30 iterations \textit{vs.} 8 for the approximant and that the latter was not regularized. The DNN reconstructions appear to be the best according to the PCC metric, which shows that, even while starting from a poor approximation, the DNN was able to outperform LT. Note that a direct comparison between the performance of the DNN on the synthetic data \textit{vs.} the experimental example is not fair because the ground truth is not known in the experiment. The ground truth used for the experimental example is an idealization from the design parameters used to fabricate the sample. We also indicate the values for reconstructions based on regularized approximant (using the same regularization parameters than the LT algorithm). In terms of PCC, there is no significant difference from the unregularized case.

\begin{table}[hb!]
	\centering
\caption{\small{Pearson correlation coefficient (PCC), expressed in percent (\textit{i.e.} PCC$\times$100), of the reconstructions in the test set with respect to the ground truths for the approximant (not regularized) and the DNN reconstructions, labeled `DNN', obtained from the unregularized approximant. We show the two cases $K=1$ and $K=8$ for the approximant calculation. The learning tomography (LT) solution is obtained with $K=30$ and are indicated on the right. The values for the DNN trained with regularized approximants are labeled `DNN reg.'. The uncertainty values indicated correspond to the standard deviation over the 50 examples of test set. For each case, the values for the synthetic (simulated) and experimental examples are indicated in separated columns `Simul.' and `Exp.' respectively. No uncertainty is given for the experimental case as it contains only one example.}}
\label{tab:pcc}
\includegraphics{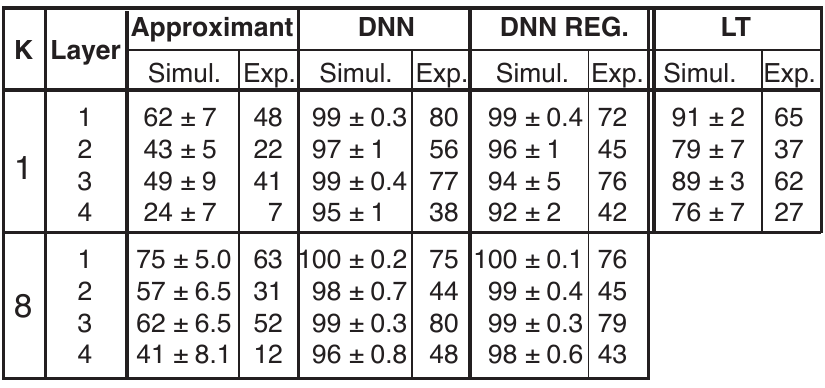}
\end{table}

\begin{figure*}[ht!]
  \begin{center}
    \includegraphics{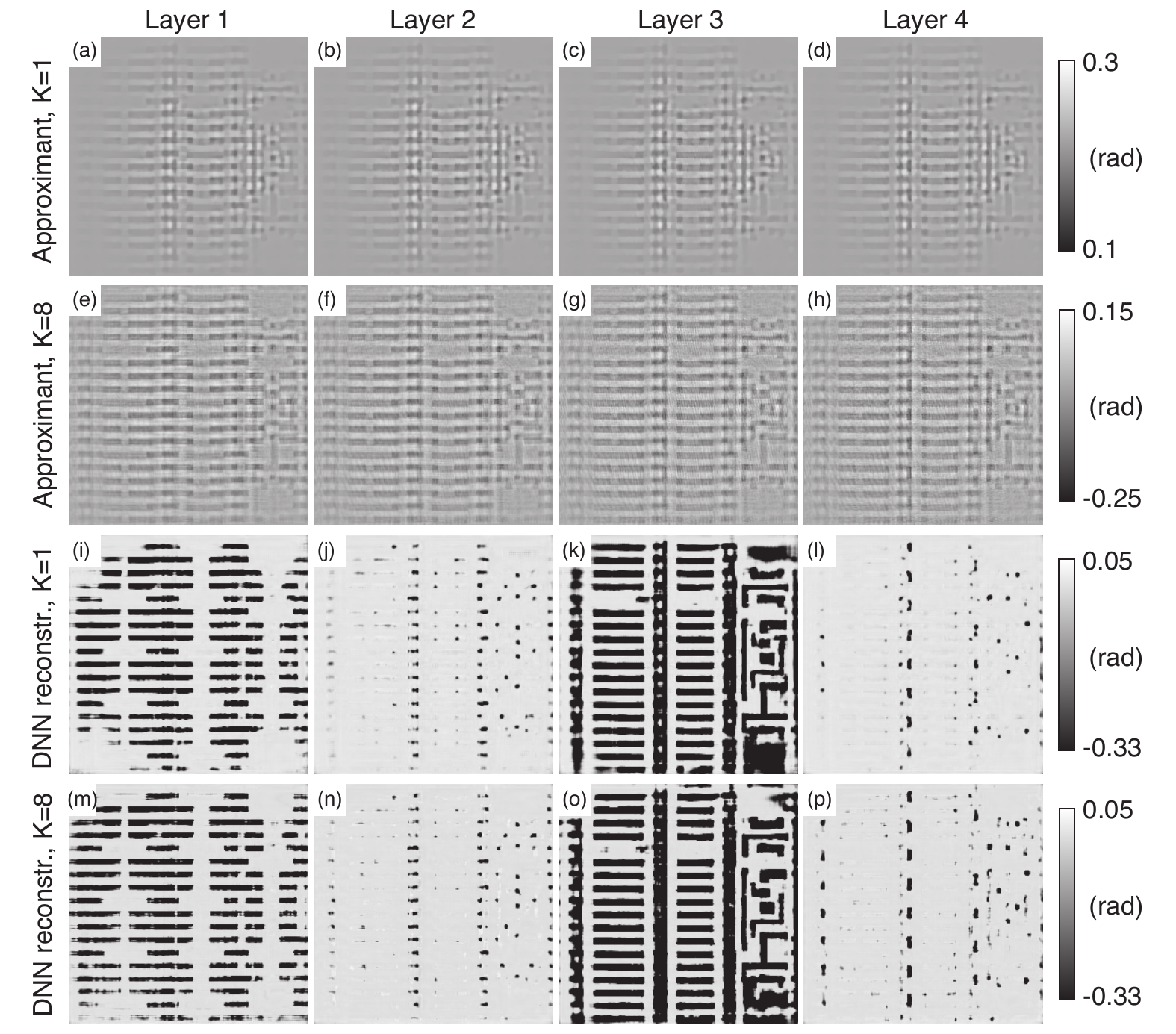}
     \caption{\small{(a) to (d) Approximants generated from the experimental measurements with $K=1$ and TV regularization with $\alpha = 0.1$. (e) to (h) Approximants obtained with $K=8$ and TV regularization with $\alpha = 0.04$. (i) to (l) Reconstructions from the DNN of each approximant a to d respectively. (m) to (p) Reconstructions from the DNN of each approximant e to h respectively.}}
	 \label{fig:reconstructions_reg}
  \end{center}
\end{figure*}

The reconstructions based on regularized approximants are shown in Fig.~\ref{fig:reconstructions_reg}. By comparing these images with the unregularized reconstructions shown in Fig.~\ref{fig:reconstructions}, and also by considering the value of the PCC in Table~\ref{tab:pcc}, we conclude that the regularization has little effect, especially on the experimental reconstructions. Moreover, the TV regularization may not operate as a favorable pre-conditioner for the DNN. The choice of the TV operator as a regularizer is arbitrary and only based on our assumption that the solution should be piecewise constant. In fact, because of the small angle range, the approximants for the different phantom layers are quite similar to each other and the regularization may cancel information that the DNN could use to discriminate between them. Layers 3 and 4 can be said to look visually better with the regularized approximant, but the situation is reversed for layer 1 and 2. Iterating more, \textit{i.e.} using $K=8$ \textit{vs.} $K=1$, yields slightly better results as can be expected intuitively, but the improvement is quite minute considering the increased computation time required to perform 7 more iterations.

In the regularized case $K=1$ only we observed instability in the behavior of the DNN for the regularized approximants. For bipolar input (approximant conataining both positive and negative values), one of the phantom layers (layer \#3) would always be reconstructed to null values. As we are using a rectified linear unit as activation function, this means that the output of one layer within the network display only negative values. By offsetting the input to the DNN (approximant) so that all values are positive we were able to remove the problem (reconstructions of Fig.~\ref{fig:reconstructions_reg}i to l). For the regularized $K=1$ case where this behavior was observed, the difference between approximant layers is the smallest, {\it i.e.} the failure may be due to the regularizer washing out the differences. 

So far, we have reconstructed the phase shift distribution $\varphi(x, y)$ associated to each layer. In fact, it is possible, with the same method, to infer the refractive index $n(x, y)$ of the sample. For a given layer, the refractive index is simply given by $n(x, y) = \varphi(x, y)/(k\Delta z)$, where $\Delta z$ is the thickness of the layer. If the layer thicknesses are not known, one would instead slice the object into layers at finer spacing to meet the applicable Nyquist criterion.

\appendix

\section{Derivation of the gradient}

This derivation follows a path similar to the derivation given in \cite{kamilov:2016}. We start from Eq.~\ref{equ:problem}:
\begin{eqnarray}\label{equ:prob2}
J &=& \frac{1}{2}\sum_{i=1}^{N_v}\|H_i(\bfu) - \bfg_i\|^2_2\\\label{equ:prob3}
  &=& \frac{1}{2}\sum_{i=1}^{N_v}\left(H_i(\bfu)^T H_i(\bfu) - 2H_i(\bfu)^T\bfg_i + \bfg_i^T\bfg_i\right).
\end{eqnarray}
\noindent The gradient of $J$ is defined as:
\begin{equation}\label{equ:grad0}
\nabla_{\bfu}J = \left[\frac{\partial J}{\partial \varphi_1}, ... , \frac{\partial J}{\partial \varphi_N}, \frac{\partial J}{\partial \alpha_1}, ... , \frac{\partial J}{\partial \alpha_N}\right]_{\bfu^{(k)}},
\end{equation}
\noindent where the object function is defined as $\bfu = \exp[j\bphi - \boldsymbol{\alpha}]$, $\bphi$ representing the phase delay and $\boldsymbol{\alpha}$ the absorption. In what follows we denote the gradient by $\nabla$ for notational simplicity. We take the gradient of Eq.~\ref{equ:prob3} and, by linearity of the derivation operation and denoting $H_i(\bfu)$ by $\bfH_i = (H_1,...,H_M)^T$, we get:
\begin{equation}\label{equ:grad1}
\nabla J = \frac{1}{2}\sum_{i=1}^{N_v}\left[\nabla (\bfH_i^T \bfH_i) - 2\nabla (\bfH_i^T\bfg_i)\right].
\end{equation}
\noindent The term $\nabla(\bfg_i^T\bfg_i)$ is absent because measurements $\bfg_i$ do not depend on the estimate $\bfu$. Then, by the following definition:
\begin{equation}\label{equ:jacobian}
\nabla \bfH =
\begin{bmatrix}
	\frac{\partial H_1}{\partial \varphi_1} & ... & \frac{\partial H_1}{\partial \varphi_N} & \frac{\partial H_1}{\partial \alpha_1} & ... & \frac{\partial H_1}{\partial \alpha_N}\\
	... & & \\
	\frac{\partial H_M}{\partial \varphi_1} & ... & \frac{\partial H_M}{\partial \varphi_N} & \frac{\partial H_M}{\partial \alpha_1} & ... & \frac{\partial H_M}{\partial \alpha_N}\\
\end{bmatrix},
\end{equation}
\noindent we get:
\begin{eqnarray}\label{equ:grad3}
\nabla J &=& \sum_{i=1}^{N_v}\left[\bfH_i^T \nabla \bfH_i - \bfg_i^T \nabla \bfH_i \right]\\\label{equ:grad4}
         &=& \sum_{i=1}^{N_v}\left[\mathbf{r}_i^T \nabla \bfH_i\right],
\end{eqnarray}
\noindent where $\mathbf{r}_i$ is the residual defined as $\mathbf{r}_i = \bfH_i - \bfg_i$. Finally, we get the expression required in Eq.~\ref{equ:grad_desc}:
\begin{equation}\label{equ:grad5}
(\nabla J)^T = \sum_{i=1}^{N_v}(\nabla \bfH_i)^T \mathbf{r}_i.
\end{equation}

\noindent In Eq.~\ref{equ:grad5}, $(\nabla \bfH_i)^T$ is a matrix of size ($2N\times M$) that is too large to be computed directly. Instead, we use a routine, described below, to calculate the vector $(\nabla \bfH_i)^T \mathbf{r}_i$ directly. We remind Eqs.~\ref{equ:forwardop0} and~\ref{equ:forwardop01} that describe the forward model where we drop the index $i$ to simplify the notation as the expression assumes the same form for each sample orientation:
\begin{eqnarray}\label{equ:forwardop}
\bfH &=& \left|\bu_{\mathrm{det}}\right|^2\\\label{equ:forwardop1}
\bu_{\mathrm{det}} &=& F_d \diag{\bfu_L}...F_{\Delta z}\diag{\bfu_2}F_{\Delta z}\diag{\bfu_1}\bu_{\mathrm{inc}}.
\end{eqnarray}

\noindent This forward operator allows for a convenient computation of the gradient by using a backpropagation scheme. We first calculate the gradient of Eq.~\ref{equ:forwardop}:

\begin{eqnarray}\label{equ:grad_derivation}
\nabla \bfH &=& \nabla|\mathbf{u}_{\mathrm{det}}|^2\\
&=& \nabla(\diag{\mathbf{u}_{\mathrm{det}}^\ast} \mathbf{u}_{\mathrm{det}})\\
&=& \diag{\mathbf{u}_{\mathrm{det}}}\nabla\mathbf{u}_{\mathrm{det}}^\ast + \diag{\mathbf{u}_{\mathrm{det}}^\ast}\nabla\mathbf{u}_{\mathrm{det}}\\
&=& 2\Re\left\{\diag{\mathbf{u}_{\mathrm{det}}^\ast}\nabla\mathbf{u}_{\mathrm{det}}\right\},
\end{eqnarray}

\noindent where the asterisk represents the complex conjugate. Thus:
\begin{equation}\label{equ:grad_derivation2}
(\nabla \bfH)^\dagger \mathbf{r} = 2\Re\left\{(\nabla\mathbf{u}_{\mathrm{det}})^\dagger\mathbf{r}'\right\}.
\end{equation}

\noindent where the dagger represents the Hermitian transpose and we have defined $\mathbf{r}' = \diag{\mathbf{u}_{\mathrm{det}}}\mathbf{r}$. Because it is not practical to compute the matrix $(\nabla\mathbf{u}_{\mathrm{det}})^\dagger$ due to its size, we use a recursive scheme to compute $(\nabla\mathbf{u}_{\mathrm{det}})^\dagger\mathbf{r}'$ directly. For that, we rewrite Eq.~\ref{equ:forwardop1} as a recursive relationship for the optical field $\bu_l$ just after layer $l$:
\begin{eqnarray}\label{equ:forwardfield1}
\bu_1 &=& \diag{\bfu_1}\bu_{\mathrm{inc}}\\\label{equ:forwardfield2}
\bu_{l} &=& \diag{\bfu_l}F_{\Delta z}\bu_{l-1}\\\label{equ:forwardfield3}
\bu_{\mathrm{det}} &=& F_d\bu_L
\end{eqnarray}

\noindent The optical field $\bu$ is thus known everywhere for a given object function $\bfu$. We then propagate the residual $\mathbf{r}'$ backward through the sample by using the same propagation relationships:
\begin{eqnarray}\label{equ:res1}
	\mathbf{r}'_L &=& F_d^\dagger \mathbf{r}'\\\label{equ:res2}
\mathbf{r}'_{l-1} &=& F_{\Delta z}^\dagger\diag{\bfu_l}^\dagger \mathbf{r}'_l
\end{eqnarray}

\noindent Note that the Fresnel operator is unitary, \textit{i.e.} $F^\dagger = F^{-1}$. We take the gradient of Eqs.~\ref{equ:forwardfield1} to~\ref{equ:forwardfield3}:
\begin{eqnarray}\label{equ:gradrec01}
\nabla\bu_{1} &=& \diag{\bu_{\mathrm{inc}}}\nabla\bfu_1\\\label{equ:gradrec02}
\nabla\bu_{l} &=& \diag{F_{\Delta z}\bu_{l-1}}\nabla\bfu_l + \diag{\bfu_l}F_{\Delta z}\nabla\bu_{l-1}\\\label{equ:gradrec03}
\nabla\bu_{\mathrm{det}} &=& F_d\nabla\bu_{L}
\end{eqnarray}

\noindent We then take the Hermitian transpose and multiply by the residual, we get:

\begin{eqnarray}
(\nabla\bu_{1})^\dagger\mathbf{r}'_1 &=& (\nabla\bfu_1)^\dagger\diag{\bu_{\mathrm{inc}}}^\dagger\mathbf{r}'_1\\\label{equ:gradrec2}
(\nabla\bu_{l})^\dagger\mathbf{r}'_l &=& (\nabla\bfu_l)^\dagger\diag{F_{\Delta z}\bu_{l-1}}^\dagger \mathbf{r}'_l+\nonumber\\
&& + (\nabla\bu_{l-1})^\dagger F_{\Delta z}^\dagger\diag{\bfu_l}^\dagger\mathbf{r}'_l\\\label{equ:gradrec3}
(\nabla\bu_{\mathrm{det}})^\dagger\mathbf{r}' &=& (\nabla\bu_{L})^\dagger F_d^\dagger\mathbf{r}'
\end{eqnarray}

\noindent We simplify the equations above by making use of Eqs.~\ref{equ:forwardfield2} and~\ref{equ:forwardfield3}:
\begin{eqnarray}\label{equ:gradrec4}
(\nabla\bu_{1})^\dagger\mathbf{r}'_1 \hspace{-0.3cm}&=&\hspace{-0.3cm} (\nabla\bfu_1)^\dagger\diag{\bu_{\mathrm{inc}}}^\dagger\mathbf{r}'_1\\\label{equ:gradrec5}
(\nabla\bu_{l})^\dagger\mathbf{r}'_l \hspace{-0.3cm}&=&\hspace{-0.3cm} (\nabla\bu_{l-1})^\dagger \mathbf{r}'_{l-1} + (\nabla\bfu_l)^\dagger\diag{F_{\Delta z}\bu_{l-1}}^\dagger\mathbf{r}'_l\\\label{equ:gradrec6}
(\nabla\bu_{\mathrm{det}})^\dagger\mathbf{r}' \hspace{-0.3cm}&=&\hspace{-0.3cm} (\nabla\bu_{L})^\dagger \mathbf{r}'_L
\end{eqnarray}

\noindent which gives us a recursive relationship to calculate the gradient of the field at each layer. Note that $(\nabla\bfu_l)^\dagger$ is a matrix of size $2N\times M$ whose entries are non-zeros only for the diagonal entries corresponding to layer $l$ because $f_i$ depends only on $\alpha_i$ and $\varphi_i$. In practice, $(\nabla\bu_{\mathrm{det}})^\dagger\mathbf{r}'$ can be built layer by layer by stacking the second term of the right hand side of Eq.~\ref{equ:gradrec5} which reads, for pure phase objects ($\boldsymbol{\alpha} = \mathbf{0}$):
\begin{eqnarray}\label{equ:gradrec7}
&& \hspace{-1.5cm}(\nabla\bfu_l)^\dagger\diag{F_{\Delta z}\bu_{l-1}}^\dagger\mathbf{r}'_l = -j\diag{e^{-j\bphi_l}}\diag{F_{\Delta z}\bu_{l-1}}^\dagger\mathbf{r}'_l\\
&=&\hspace{-0.3cm} -j\diag{e^{-j\bphi_l}}\diag{e^{j\bphi_l}}\diag{\bu_{l}^\ast}\mathbf{r}'_l\\
&=&\hspace{-0.3cm} -j\diag{\bu_{l}^\ast}\mathbf{r}'_l,
\end{eqnarray}
\noindent where we have used Eq.~\ref{equ:forwardfield2}. Finally, according to Eq.~\ref{equ:grad_derivation2}, we obtain layer $l$ of $(\nabla \bfH)^\dagger \mathbf{r}$:
\begin{equation}
(\nabla \bfH)^\dagger \mathbf{r}|_{\mathrm{layer}\ l} = 2\Im\left\{\diag{\bu_{l}^\ast}\mathbf{r}'_l\right\}.
\end{equation}
\noindent where $\Im$ denotes the imaginary part.

\matmethods{The experimental apparatus is shown in Fig.~\ref{fig:setup}(a). The light source is a CW He-Ne laser at $\SI{632.8}{\nano\meter}$ that is spatially filtered, expanded and collimated into a quasi plane wave with an Airy disk intensity profile of 33mm in diameter. The sample is mounted on a 2-axis rotation stage rotating along the x and y axes. The sample middle plane is imaged using a demagnifying telescope ($\times 0.50$) lens system in order to enhance the effect of diffraction and increase the field of view on the detector. The detector is an EM-CCD (QImaging Rolera EM-C2) with a 1004$\times$1002 array of $8\times\SI{8}{\micro\meter}$ pixels. In order to simulate the diffraction occurring in an X-ray measurement, the detector is defocused by a distance $\Delta z = 58$mm from the image plane, which corresponds to Fresnel numbers ranging from 0.7 to 5.5 for the different object features.

The sample corresponds to a $10^4\times$ scale up of a real IC design. The original IC comprises 13 layers, including the doped layers. We only kept layers 5 to 8 from the original design (relabeled here 1 to 4) shown in Fig.~\ref{fig:setup}(c), that contain copper patterns that would induce a significant phase delay in the X-ray regime. The four glass plates corresponding to the IC layers were cut in a $\SI{500}{\micro\meter}$ thick fused silica wafer and $575\pm 5$nm deep patterns (measured after fabrication with a Bruker DekTak XT stylus profilometer) were obtained by wet etching. In order to control the phase contrast and reduce parasitic reflections between the layers, we used an immersion oil (see Fig.~\ref{fig:setup}(b)) with a refractive index $n_D = 1.400\pm 0.0002$ at 25$^\circ$C from Cargille-Sacher Laboratories. According to the manufacturer, the refractive index of the oil is $n_{\mathrm{oil}}=1.4005\pm 0.0002$ at $\SI{632.8}{\nano\meter}$ and 20$^\circ$C. The refractive index of fused silica is $n_{\mathrm{glass}} = 1.457$ at $\SI{632.8}{\nano\meter}$ and 20$^\circ$C\ \cite{malitson:1965}, which gives a contrast of $\Delta n = 0.0565\pm 0.0005$. The corresponding phase shift for a single pattern is then $\Delta \varphi = k d \Delta n = 0.323\pm 0.006$ rad.

The sample layers are fabricated on double-sided polished $\SI{150}{\milli\meter}$ diameter and $\SI{500}{\micro\meter}$ thick fused silica wafers. A $\SI{1}{\micro\meter}$ thick positive tone resist (Megaposit SPR700) is spin-coated at 3500 rpm on both sides of the wafer and soft-baked at 95$^\circ$C for 30 mins in a convection oven. The backside was also coated for protection from the forthcoming wet-etch. Scaled versions of IC designs in GDSII format are then patterned directly using a maskless aligner (MLA150, Heidelberg Instruments, Heidelberg, Germany) with a 405nm laser and developed using an alkaline developer (Shipley Microposit MF CD-26) for 45s followed by de-ionized (DI) water rinse and N$_2$ drying. A hard-bake at 120$^\circ$C for 30 mins is carried out to stabilize the patterned features. A short descum of 2 mins at 1000 W and 0.1 Torr O$_2$ pressure in a barrel asher is also performed to remove any resist residue. The wafers are subsequently etched for 7 mins at a rate of $\sim\SI{80}{\nano\meter}$/min in buffered oxide etch. The resist is then stripped from the wafer by a long ash (1 hour) followed by a Piranha clean (3:1 H$_2$SO$_4$:H$_2$O$_2$), DI water rinse and N$_2$ drying. Finally, the wafers are diced into squares of $\SI{50}{\milli\meter}$ by $\SI{50}{\milli\meter}$ and cleaned again with Piranha, DI water rinse and N$_2$ drying.

}

\showmatmethods{} 

\acknow{This work was supported by the Intelligence Advanced Research Projects Activity (IARPA) No. FA8650-17-C-9113.}

\showacknow{} 

\bibliography{DNNTomography}

\end{document}